\documentclass[twocolumn]{aastex631}

\usepackage{gensymb}
\usepackage{amsmath}
\usepackage{amsfonts}
\usepackage{graphicx}
\usepackage{hyperref}
\usepackage{float}
\usepackage{multirow}
\usepackage[mathlines]{lineno}

\begin{document}

\title{Supersonic Motion in the Driving Region of M82}

\author{Skylar Grayson}
\affiliation{School of Earth and Space Exploration, Arizona State University, P.O. Box 876004, Tempe, AZ 85287, USA}

\author{Evan Scannapieco}
\affiliation{School of Earth and Space Exploration, Arizona State University, P.O. Box 876004, Tempe, AZ 85287, USA}

\author{Philipp Grete}
\affiliation{Hamburger Sternwarte, University of Hamburg, Gojenbergsweg 112, 21029 Hamburg, Germany}

\author{Erin Boettcher}
\affiliation{Department of Astronomy, University of Maryland, College Park, MD 20742, USA}
\affiliation{X-ray Astrophysics Laboratory, NASA/GSFC, Greenbelt, MD 20771, USA}
\affiliation{Center for Research and Exploration in Space Science and Technology, NASA / GSFC (CRESST II), Greenbelt, MD 20771, USA}

\author{Marcus Br{\"u}ggen}
\affiliation{Hamburger Sternwarte, University of Hamburg, Gojenbergsweg 112, 21029 Hamburg, Germany}

\author{Edmund Hodges-Kluck}
\affiliation{NASA Goddard Space Flight Center, Greenbelt, MD 20771, USA}

\author{John ZuHone}
\affiliation{Center for Astrophysics, Harvard \& Smithsonian}

\author{Yutaka Fujita}
\affiliation{Department of Physics, Tokyo Metropolitan University, Tokyo 192-0397, Japan}

\submitjournal{The Astrophysical Journal}

\begin{abstract}

The prototypical starburst galaxy M82 is host to an expansive, multiphase outflow whose driving mechanism is not fully understood. Longstanding models suggest that energy and mass injection from supernova into the hottest phase of the galactic wind could drive the cooler phases, but validating these models has been difficult due to the lack of constraints on the hot wind energetics. The high-resolution spectral capabilities of XRISM have generated the tightest constraints to date on the temperatures of the hot wind, as well as the first direct measurement of its velocity dispersion. In this work, we use these new observational constraints to test a model of a supernova-driven free wind. We generate a suite of highly idealized hydrodynamic simulations varying the energy and mass loading of the starburst and construct mock spectra to compare against the XRISM results. We find that the observed velocity dispersion is impossible to replicate using our free-wind model alone, and extra broadening is required to fit the spectrum. We interpret this broadening to be due not to bulk outflow, but rather to smaller scale non-thermal motions in the driving region of the starburst. This implies supersonic motion (Mach 1.71-3.14) of the hot gas in the central region of the galaxy. As supersonic motions are unexpected, it is possible that a significant amount of the energy that should go into heating the gas is instead going towards other sources such as amplifying magnetic fields and driving cosmic rays. 

\end{abstract}


\section{Introduction}
\label{sec:intro}

Galaxies undergoing rapid star-formation in the nearby universe can generate multikiloparsec-scale outflows that inject metals and energy into the circumgalactic medium \citep[see][for a review]{thompsonandheckman}. The starburst galaxy M82 provides an excellent case study of these processes, with an extended outflow visible across the electromagnetic spectrum \citep{strickland_2009}. This superwind has been observed to contain dust \citep{Engelbracht2006, Bolatto2024} as well as molecular \citep{walter2002, Veilleux2009, bolatto2013, leroy2015}, neutral \citep{Rupke2005, martini2018}, warm ionized \citep{McCarthy1987, Shopbell1998, Westmoquette2009a, Westmoquette2009b}, and hot ($T>10^7$ K) gas \citep{Strickland2004a,Strickland2004b,lopez2020}. The complex interactions between these phases remain poorly understood, with one of the biggest open questions being the driving mechanism behind this multiphase outflow.

Longstanding models of starbursts suggest that the outflow could be driven by its hottest (T $\approx 10^7$ K) phase, powered by the energy and mass injection of supernovae and stellar winds in the core of the galaxy. The foundational model developed in \cite{CC} (hereafter CC) provides an analytic solution for a free wind driven by a constant mass and energy loading rate within a spherical driving region.

The one-dimensional solution is dependent on three parameters: the energy injection rate $\dot{E}$, the mass injection rate $\dot{M}$, and the driving region radius $R_*$. This yields a solution where 
\begin{equation}
    \left( \frac{3\gamma+1/M^2}{1+3\gamma}\right)^{-\frac{3\gamma+1}{5\gamma+1}} \left( \frac{\gamma-1+2/M^2}{1+\gamma}\right)^{\frac{\gamma+1}{10\gamma+2}}=\frac{r}{R_*},
\end{equation}
when $r<R_*$. When $r>R_*$, the solution takes the form
\begin{equation}
    M^{\frac{2}{\gamma-1}}\left( \frac{\gamma-1+2/M^2}{1+\gamma}\right)^{\frac{\gamma+1}{2\gamma-2}}=\left( \frac{r}{R_*}\right)^2.
\end{equation}
In these equations $\gamma$ is the adiabatic index, $M$ is the Mach number, and $r$ is the radius. We can find the density $\rho$ by integrating these equations, yielding 
\begin{equation}
    \rho = \frac{q_Mr}{3v},
\end{equation}
when $r<R_*$ and
\begin{equation}
    \rho = \frac{q_MR_*^3}{3vr^2},
\end{equation}
when $r>R_*$. $q_M =\dot{M}/V_*$ for the starburst volume $V_*=4\pi R_*^3/3$. The velocity of the outflow, $v$, is given by $v=Mc_s$ where $c_s=\sqrt{\gamma P/\rho}$ and $P$ is the pressure.

The thermal and ram pressure of this hot phase could then drive cooler phases \citep[e.g.][]{Scannapieco2015, zhang2017}, resulting in the complex structure observed today. However, validating this sort of free-wind model has proven difficult, largely due to the lack of strong constraints on the energetics of the hot wind. 

The X-ray Imaging and Spectroscopy Mission (XRISM) provides the capabilities needed to place constraints on the temperatures and velocities of this material. The Resolve microcalorimeter on XRISM provides a spectral resolution of 4.5 eV in energies from 1.8-10 keV \citep{xrism2021}, allowing for the first measurement of the hot wind velocity dispersion. M82 was observed during the Performance Verification phase of XRISM in 2024, with a total cleaned exposure of 207 ks. An inital analysis of the XRISM data is presented in \cite{Boettcher}, setting new constraints on the temperature ($T = 2.3^{+0.5}_{-0.2} \times 10^7$ K) and measuring the velocity dispersion ($\sigma_{\rm LOS} = 595^{+464}_{-128}$ km/s) of the hot wind in M82 for the first time. These measurements offer a new opportunity to test and verify models of starburst-driven winds. 

Preliminary comparisons against a CC model in \cite{Boettcher} found that a free-wind model  matched to the observed temperature underpredicts the velocity dispersion compared to XRISM observations. In this work, we explore the implications of the higher-than-expected velocity dispersion by generating a suite of hydrodynamic free-wind simulations, creating simulated spectra, and comparing with the XRISM results. This allows us to explore a wide range of input energy and mass rates, as well as different outflow geometries, in order to test the robustness of a CC-like model in the face of the novel XRISM constraints.  

The structure of this work is as follows. In Section \ref{sec:methods}, we outline the methodology for generating the simulations and constructing and fitting mock spectra. In Section \ref{sec:res}, we share the best-fit constraints on energy and mass loading and explore the velocity structure of the hot wind fluid. In Section \ref{sec:disc}, we explore what these results mean in the context of starburst-driven outflow models. 

\section{Methodology}
\label{sec:methods}
\subsection{AthenaPK}
The simulations were conducted with the open source, performance portable (magneto)hydrodynamics code \texttt{AthenaPK}\footnote{\texttt{AthenaPK} is available and maintained at \url{https://github.com/parthenon-hpc-lab/athenapk} and commit \texttt{4f9ee52} was used for the simulations presented in this paper.}
based on the adaptive mesh refinement library \texttt{Parthenon} \citep{grete2022} and Kokkos \citep{trott2022}. We employ an overall second-order accurate scheme consisting of Van-Leer type, predictor-corrector integrator, HLLC Riemann solver, and piecewise-linear reconstruction to solve the compressible Euler equations with an ideal equation of state with an adiabatic index of $\gamma = 5/3$. Internal, embedded boundaries have been implemented using a simple, first-order method: boundaries are always aligned with cell interfaces at which reflective conditions are enforced.

We set up our simulations in a (4 kpc)$^3$ box with $256^3$ cells, giving a spatial resolution of $\sim$16 pc. We ran a convergence test with $512^3$ cells, finding slight differences ($<2\%$) in the density and temperature in inner 500 pc due to the different rendering of the geometry. These differences have no impact on the overall results of the paper, and so we proceeded with the lower resolution model to allow for a larger sample of simulations. We model a starburst geometry consisting of a spherical driving region with hard boundaries, except along a biconical outflow parameterized by a fixed opening angle. We add mass and thermal energy uniformly to the fluid within the spherical driving region, parameterized by total injection rates $\dot{M}$ and $\dot{E}$. The initial density in the gas is $1 \times 10^{-27}$ g cm$^{-3}$ and initial gas temperatures are set at $10^7$ K, but we find the results are not dependent on these initial conditions. We do not include a gravitational potential. Each run is carried out until reaching a steady state within 1 kpc, approximately 20 Myr. It is worth noting that it takes such a long time to reach equilibrium due to the geometry of the model and effects that arise due to the construction of the simulation with a hard boundary around the driving region.

\begin{figure}
\centering
\includegraphics[width=\linewidth]{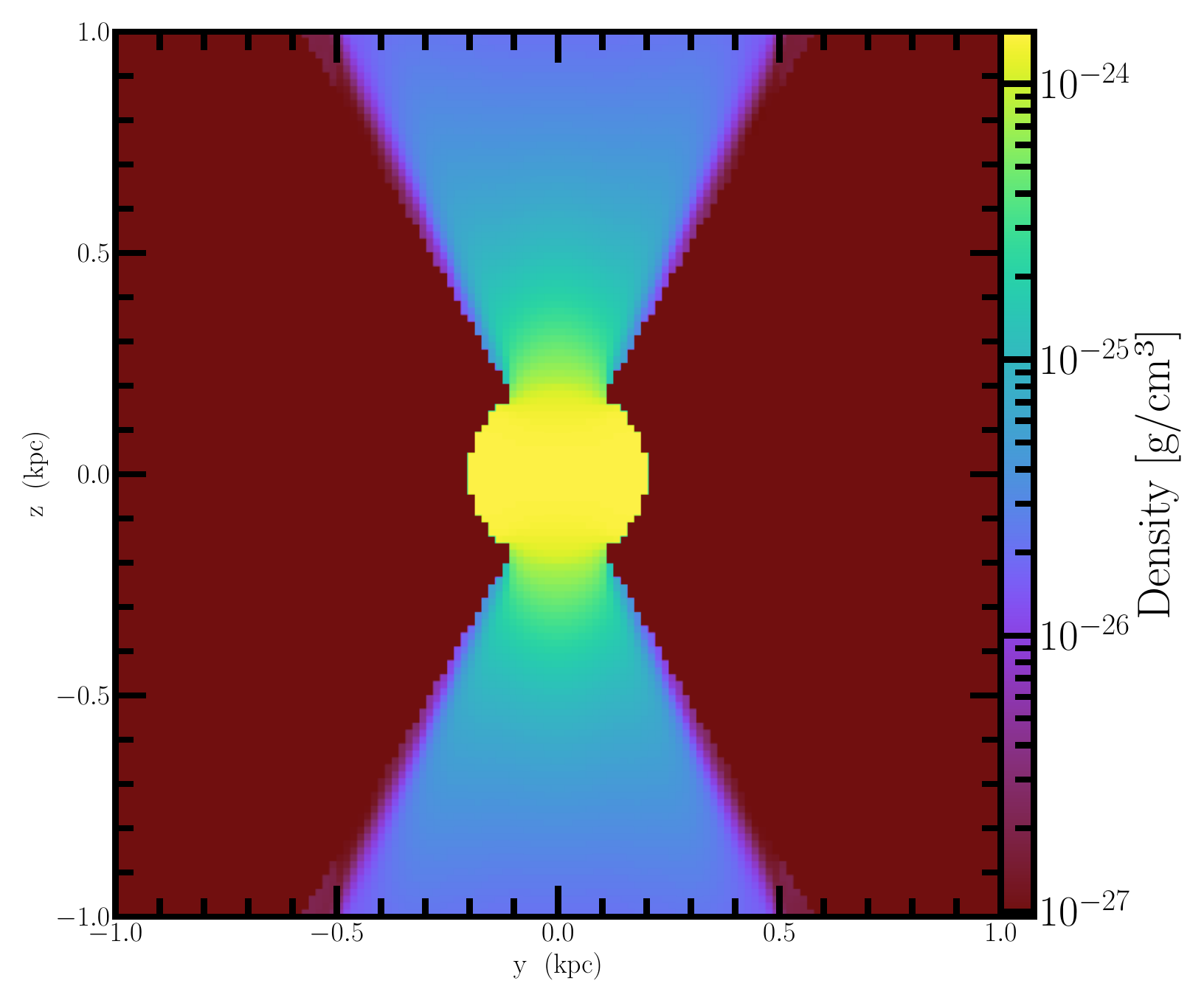}
\caption{Slice plot of the 3D \texttt{AthenaPK} output showing the density structure for a run with a half opening angle of 30$\degree$.}
\label{fig:slice}
\end{figure}

We explore three different geometries within \texttt{AthenaPK}. All three use a spherical driving region with a radius of 200 pc. This was chosen to match the volume of the driving region identified in \cite{Boettcher}, which was motivated by Chandra observations of Fe XXV emission within an ellipsoid in the center of M82 \citep{Strickland2007,Iwasawa2021}. We chose three half opening angles: a fiducial model of 30 degrees, and  models with 15 and 60 degrees that approximately span the range of values seen for the outflow angle in the literature. Figure \ref{fig:slice} shows a density slice of a completed simulation with a half opening angle of 30$\degree$, demonstrating the geometry of our fiducial model.  

For each geometry, we span a unique range of $\dot{E}$ and $\dot{M}$. To efficiently sample the parameter space, we used Latin hypercube sampling (LHS), selecting unique parameters for each model. We first performed LHS sampling over a broad region, then used the results to generate another set of models in a more refined parameter space. For the fiducial geometry, this consisted of an initial 500 runs spanning $\dot{E}= 0.05-10 \times 10^{41}$ erg/s and $\dot{M}= 0.1-1$ $M_\odot$/yr, then 400 runs with  $\dot{E}= 0.2-0.3 \times 10^{41}$ erg/s and $\dot{M}= 0.4-0.6$ $M_\odot$/yr. For the other opening angles, we did not attempt to conduct a complete statistical analysis, as the purpose of exploring other geometries is to set approximate lower and upper limits on the energy and mass loading for a range of reasonable opening angles. Thus, we determined the ratio of energy to mass loading that will give a temperature of 2 keV in the center of the driving region and ran 50 simulations for each geometry that scale the energy and mass, keeping the ratio constant. Table \ref{tab:parameter_space} shows the parameter space explored for each geometry.

\begin{table*}
\centering
\caption{Summary of the simulation parameter space explored in this work. Note that $\sigma_{\rm LOS}$ is a post-processing parameter, as described in Section \ref{sec:velbroadening}. The $\dot{E}$ and $\dot{M}$ parameters were explored over 900 runs using LHS sampling for the fiducial model. For the narrow and wide geometries the ratio of $\dot{E}/\dot{M}$ was kept fixed, and we ran 50 simulations per geometry. The velocity parameter space was sampled on a uniform grid of 200 values.}
\label{tab:parameter_space}
\begin{tabular}{cccccc}
\hline
Model & Half Opening Angle & Driving Region Radius & $\dot{E}$ Values & $\dot{M}$ Values & $\sigma_{\rm LOS}$ Values \\
 & ($^\circ$) & (pc) & (erg/s) & ($M_\odot$/yr) & (km/s) \\
\hline
Fiducial & 30 & 200 & $0.05-10 \times 10^{41}$  & 0.1-1 & 10-3000\\
Narrow & 15 & 200 & $0.05-1.5 \times 10^{41}$ & 0.01-0.3 & 10-3000 \\
Wide & 60 & 200 & $6.6-9.1 \times 10^{41}$ & 1.3-1.8 & 10-3000 \\
\hline
\end{tabular}
\end{table*}

\subsection{Generating Mock Spectra}

\begin{figure}
\centering
\includegraphics[width=\linewidth]{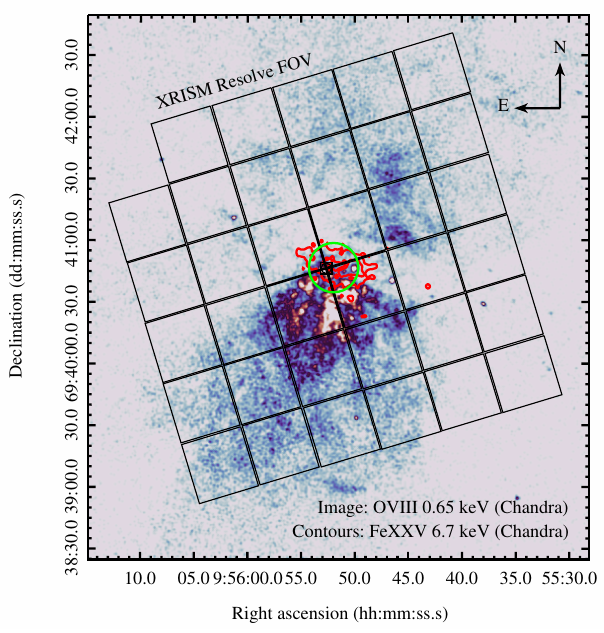}
\caption{Chandra data showing the spatial distribution of hard and soft X-ray emission relative to the XRISM Resolve pointing (black grid). The image shows OVIII 0.65 keV  emission which extends out past 2 kpc in the M82 wind. In red contours are shown the Fe XXV 6.7 keV data with point sources removed. We see that this hard X-ray emission, which is the focus of the analysis in this paper, is highly centralized. The green circle represents the region with a radius of 200 pc following our model geometry.}
\label{fig:chandra}
\end{figure}

We generated mock spectra from our simulations using the Python packages \texttt{pyXSIM} and \texttt{SOXS}. We used \texttt{pyXSIM} \citep{zuhone2016} to construct a list of photons assuming collisional ionization equilibrium via Monte-Carlo sampling of the Astrophysical Plasma Emission Code \citep{Smith2001}. We generate photons from within a 800 pc radius sphere, ensuring we are using material that has reached equilibrium and encapsulating all the hard X-ray emission. As XRISM Resolve has a large 3 kpc square field of view, we do not need to worry about any spatial offsets that might mean the driving region is not captured in the observations. Figure \ref{fig:chandra} shows the XRISM FOV overlaid over Chandra ACIS-S narrowband data. The contours show the narrowband emission around the Fe XXV 6.7 keV line, demonstrating this hard X-ray emission is highly centralized and we can confidently focus our analysis on the inner region. 

The initial photon sample was generated in 3D and then projected along a line of sight onto a sky plane, incorporating Doppler shifting due to the velocities of the fluid elements. We considered a line of sight corresponding to a viewing angle of 80 degrees relative to the disk, accounting for M82's nearly edge-on orientation relative to us.  We also placed our observations at a redshift of 0.000677 \citep{freedman1994}.

We then used \texttt{SOXS}\footnote{\url{https://hea-www.cfa.harvard.edu/soxs/}}, or the Simulated Observations of X-ray Sources package, to create mock XRISM-Resolve spectra. We used the XRISM instrument package built into \texttt{SOXS}, although we substituted the RMF file used in the analysis of \cite{Boettcher} in order to enable accurate comparisons with the observations. We “observed” for 207 ks, the same cleaned exposure time as the XRISM PV observations, and we did not introduce Poisson (counting) noise so as to produce a clean ``model'' spectrum. Additionally, the instrument and sky background were sufficiently low compared to the source that we safely neglected these terms. However, there were several additional emission components from M82 that we needed to account for in order to robustly compare the model to the data.

The analysis conducted in \cite{Boettcher} found that the spectra were best fit by a two-temperature model: the hot (kT $\approx 2$ keV) phase that we replicated in our simulations and a cooler component with kT $\approx 0.72$ keV. Additionally, the observed spectrum included emission from unresolved X-ray binaries. In order to compare our spectra to the XRISM observations, we included the best-fit models of these two components from \cite{Boettcher}. The X-ray binary emission was modeled as an absorbed power law with a hydrogen column density of $N_H = 2.3 \times 10^{22}$ cm$^{-2}$, a photon index of 1.8, and a normalization factor of $6.6 \times 10^{-3}$. We also included the kT $\approx 0.72$ keV component, but we found that it did not contribute significantly to our lines of interest, and removing it from the total spectra did not affect our overall conclusions. Both the XRB and 0.72 keV components were directly added to the mock spectra from the simulation, creating a complete model that we compared against the observations while fitting for the simulation input parameters.

As we were not interested in testing the fits for all components, we focused only on two spectral regions that were dominated by emission lines from the hot phase: S XVI at 2.5–3 keV and the Fe XXV and Fe XXVI lines at 6.5–7 keV. The non-detection of Fe XXVI set a strong upper limit on the temperature of the hot fluid, while the widths of the S XVI and Fe XXV lines governed the velocity dispersion fit. By focusing on these two line regions, we ensured that our fits were dominated by the hot component without contamination from cooler phases. We calculated C-stat \citep{humphrey2009} from the fits in both bins and used this statistic to constrain the $\dot{M}$ and $\dot{E}$ parameters.

\subsection{Velocity Broadening}\label{sec:velbroadening}

Preliminary comparisons with a CC model in \cite{Boettcher} found that the analytic solution for a free-wind model underpredicted the velocity within the driving region relative to the best fits from XRISM observations. This result was verified in our simulations, as across the range of $\dot{E}$, $\dot{M}$, and opening angles we explored, we systematically underpredicted the velocity dispersion relative to the XRISM spectrum. We thus introduced an additional step to our modeling process. We convolved the simulated spectra of the hot phase with a Gaussian profile. We assume the width of the Gaussian is determined by the line-of-sight velocity dispersion $\sigma_{\rm LOS}$, and we fit $\sigma_{\rm LOS}$ simultaneously with $\dot{E}$ and $\dot{M}$ against the XRISM observations. By assuming a broadening driven by additional velocity dispersions, we are attempting to account for small-scale motions not captured in our simple wind model. We discuss the limitations and implications of this approach in Section \ref{sec:disc_vel}.

\section{Results}
\label{sec:res}

\begin{figure}
\centering
\includegraphics[width=\linewidth]{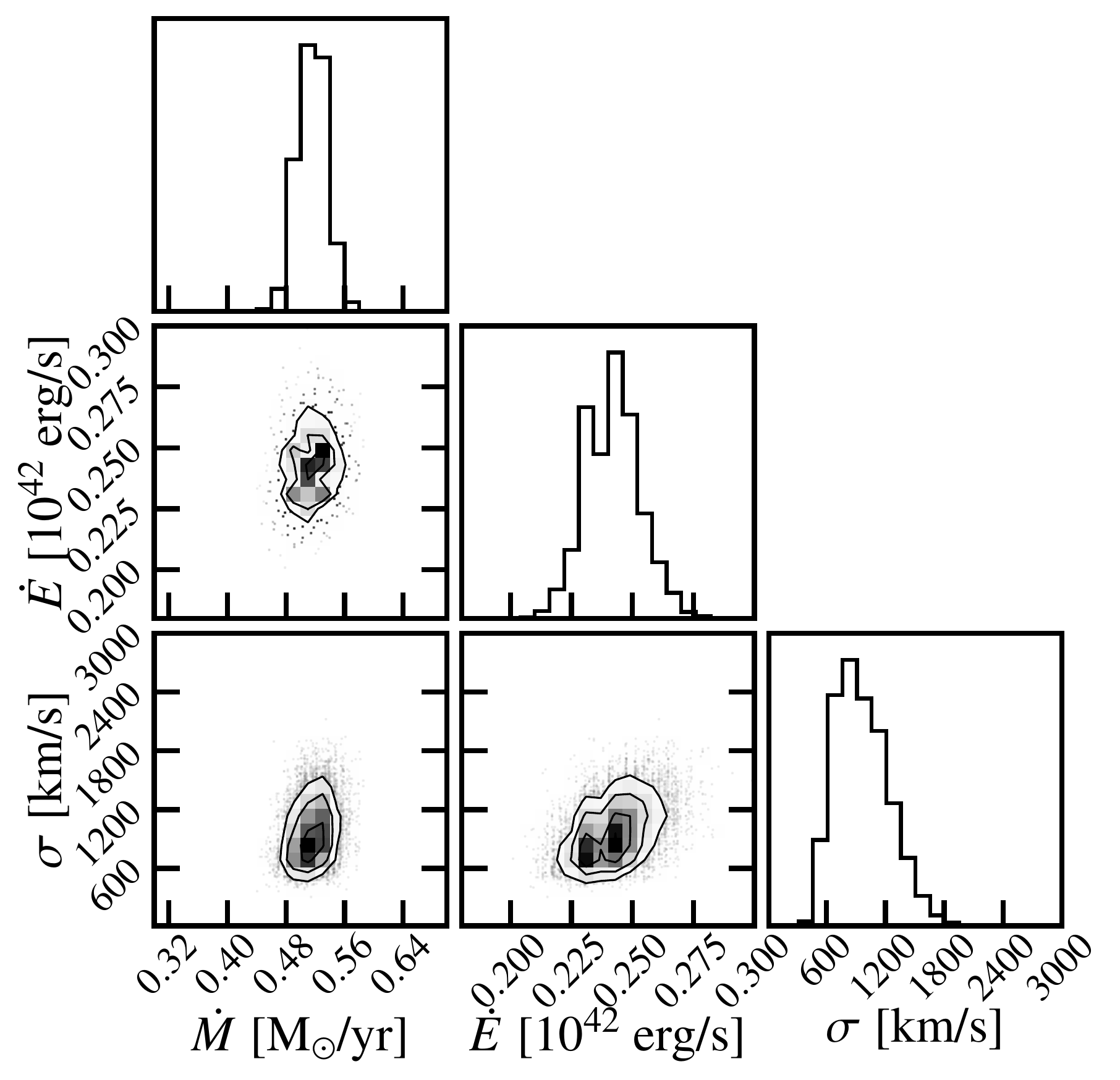}
\caption{Corner plot showing the joint constraints on $\dot{M}$, $\dot{E}$, and $\sigma_{\rm LOS}$ from runs with $\theta = 30$. Contours correspond to the 0.5 sigma-interval confidence regions for the 2D histograms. }
\label{fig:corner}
\end{figure}

Here we present the best fit constraints on $\dot{M}$, $\dot{E}$, and $\sigma_{\rm LOS}$ for our fiducial geometry model that has a driving region radius of 200 pc and a half opening angle of 30 degrees. We ran 900 simulations using two series of LHS sampling of the parameter ranges shown in Table \ref{tab:parameter_space}.
Figure \ref{fig:corner} shows the constraints on these parameters generated by fitting the Fe XXV and S XVI regions. We determine the best fit models to have $\dot{M}$ = 0.52$^{+0.020}_{-0.021}$ $M_\odot$/yr, $\dot{E}$ = 2.4$^{+0.1}_{-0.09}\times 10^{41}$ erg/s and $\sigma_{\rm LOS}$ = 825$^{+420}_{-150}$ km/s. The best fit model has a C-stat of 1555.2 for 1573 bins. Below, we explore the characteristics of the best-fit model.

\begin{figure*}
\centering
\includegraphics[width=\linewidth]{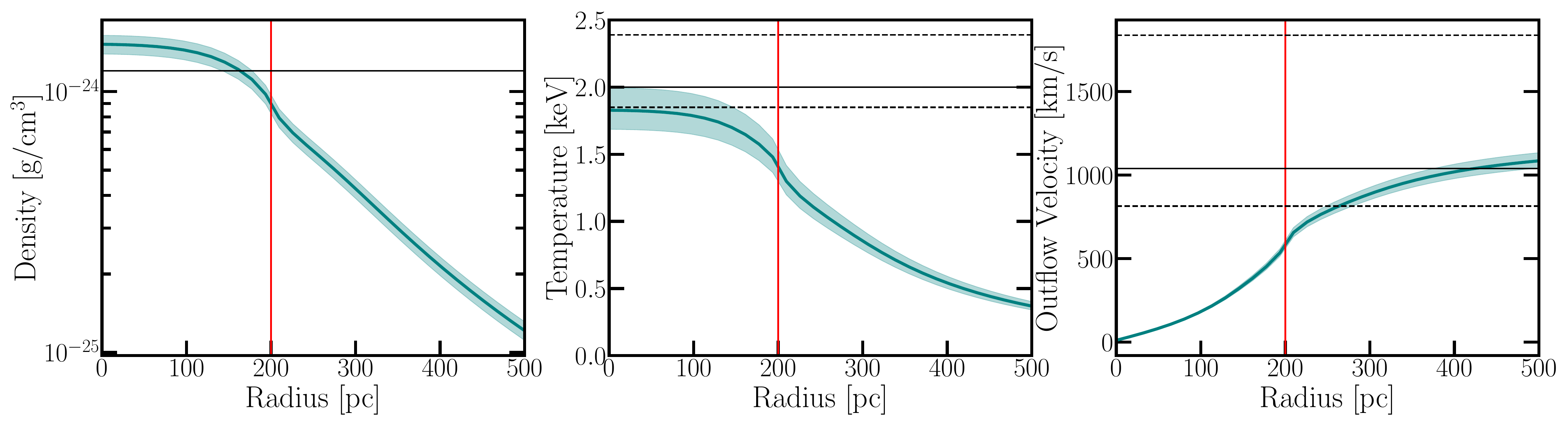}
\caption{Radial profiles (calculated along a line down the barrel of the outflow cone) of density, temperature, and velocity for a model with $\theta = 30$, $\dot{M}$ = 0.52 $M_\odot$/yr and $\dot{E}=2.4 \times 10^{41}$ erg/s. Shaded regions show the 16th and 84th percentiles on the best fits. Red line shows the radius of the spherical driving region, and the horizontal black lines in the middle and right panel show the best fits from \cite{Boettcher}, with dashed representing 1$\sigma$ errors. The velocity panel shows the $v_{RMS}$ values from \cite{Boettcher}, calculated as $\sqrt{3}\sigma_{LOS, obs}$, which provides a more accurate comparison to the velocity profiles.}
\label{fig:1dprofiles}
\end{figure*}

We can use the simulations to generate profiles of gas conditions in the wind fluid, as shown in Figure \ref{fig:1dprofiles}. Black horizontal lines represent the temperatures, densities, and velocities from \cite{Boettcher}. The velocity shown is $v_{RMS}$, calculated as $\sqrt{3}\sigma_{LOS, obs}$ in order to correct for projection effects and provide the most robust comparison to the 1D outflow velocity shown in teal. Similar to the comparisons with the analytic CC solution in \cite{Boettcher}, we find strong alignment with the measured temperature and density within the driving region, highlighting some success of this model. In the rightmost panel, we see the velocity dispersion as predicted directly from the hydro simulations.  Here, the discrepancy is apparent. While velocities can reach the observed values near the edge of the driving region, as this radius also corresponds with a steep drop-off in density, a majority of the emission in a free-wind model comes from slower-moving material than what was observed with XRISM.

\begin{figure*}
\centering
\includegraphics[width=\textwidth]{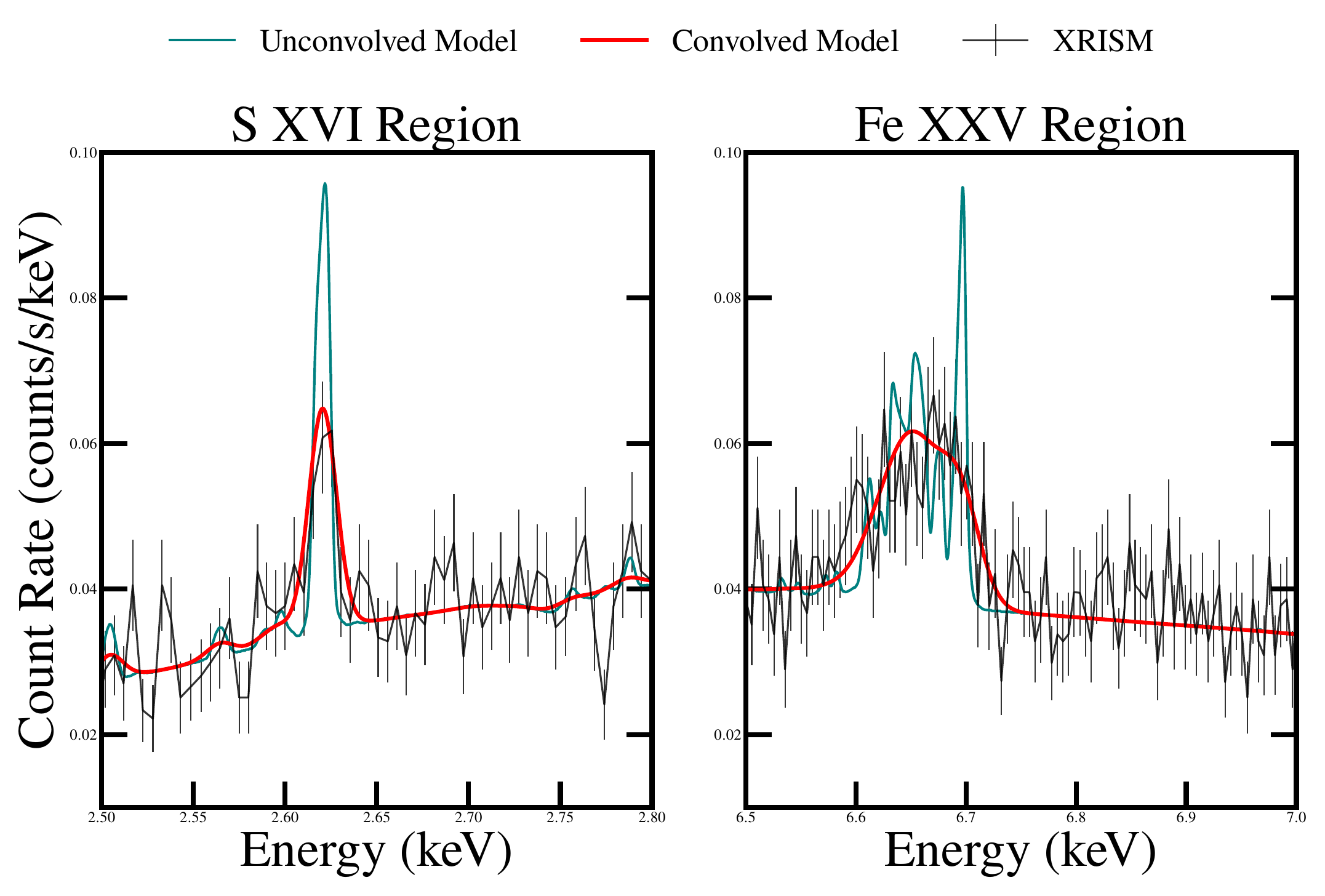}
\caption{Best fit models around the S XVI (left) and Fe XXV (right) regions. Teal shows the spectra generated from the AthenaPK simulation with $\theta = 30$, $\dot{M}$ = 0.52 $M_\odot$/yr and $\dot{E}=2.4 \times 10^{41}$ erg/s, and red shows that model convolved with a Gaussian with $\sigma_{\rm LOS}$ = 825 km/s. Both red and teal curves also include a power law and 0.7 keV model component as described in Section \ref{sec:methods}. Black is the binned Resolve spectrum, as cleaned and analyzed in \cite{Boettcher}. }
\label{fig:spectral_fits}
\end{figure*}

This is further highlighted in Figure \ref{fig:spectral_fits}, which shows the best fit spectra in the S XVI and Fe XXV regions compared against the binned XRISM Resolve observations. The teal curve, which shows the spectra generated directly from the simulation, highlights how much lower the predicted velocity dispersion is from what is observed. However, as seen with the red curve, once further velocity broadening is added to the model, we are able to better fit the observed spectra. 

\subsection{Geometric Effects}

So far, we have focused on one particular geometry for the M82 starburst, motivated by Chandra and XRISM observations. However, the geometry of the driving region remains fairly unconstrained, in particular, the opening angle of the outflow. Various works have adopted opening angles ranging from 10-80 degrees \citep[e.g.][]{heckman90,shopbell98, leroy2015, Boettcher}. In order to account for this uncertainty, we explore doubling and halving the half opening angle of our simulations to 15 and 60 degrees. We ran 50 simulations per geometry keeping $\dot{E}/\dot{M}$ fixed in order to set approximate limits for a range of reasonable opening angles.

We find the model that best matches the data for a 15 degree half opening angle has $\dot{M}=0.11$ $M_\odot$/yr and $\dot{E}$ = 0.53$\times 10^{41}$ erg/s, with a velocity broadening of $\sigma_{\rm LOS} = 765$ km/s. For a 60-degree half opening angle, we get a best-fit model of $\dot{M}=1.6$ $M_\odot$/yr and $\dot{E}$ = 8$\times 10^{41}$ erg/s, with a velocity broadening of $\sigma_{\rm LOS} = 825$ km/s.

\section{Discussion}
\label{sec:disc}

\subsection{Wind Energetics}
\label{sec:disc_energy}

The constraints on mass and energy loading found in this work have significant implications for our understanding of how galactic outflows are launched. In \cite{Boettcher}, the total energy of the hot wind was derived assuming the outflow can escape over 45$\%$ of the surface of the driving region, finding an energy outflow rate of $4.2^{+6.7}_{-2.0} \times 10^{42}$ erg/s and a mass outflow of $7_{-2}^{+3}$ M$_\odot$/yr. These values are significantly higher than our best fit fiducial model, which found $\dot{M}$ = 0.52 $M_\odot$/yr and $\dot{E}=2.4 \times 10^{41}$ erg/s. 

This discrepancy can be largely attributed to the geometric differences, as our fiducial model nozzled the wind to only escape over $\approx$14$\%$ of the surface. However, even our 60-degree half opening angle model, which corresponds to a similar escape area to that used in \cite{Boettcher}, only increases the best fits to $\dot{M}=1.6$ M$_\odot$/yr and $\dot{E}$ = 8$\times 10^{41}$ erg/s, still far below the \cite{Boettcher} values. Thus we see that free wind models with any opening angle lead to smaller assumed $\dot{M}$ and $\dot{E}$. 

This discrepancy is important to understand when determining the ability of the hot phase of the wind to power the rest of the outflow. As calculated in \cite{Boettcher}, the combined $H_2$, HI, HII, and soft X-ray observations suggest a total power in the cooler wind phases of $\dot{E}_{cool} = 2.6_{-1.2}^{+5.9} \times 10^{42}$ erg/s. This is above the energy of even our most generous geometric model, which could suggest that the thermal pressure of the hot phase alone is not enough to power the whole outflow. 

There are three important things to note here. First, the best-fit energy input is very dependent on a geometry that is highly unconstrained. This work demonstrates the extent to which the wind nozzling impacts the needed energy loading, as larger opening angles require larger $\dot{E}$ and $\dot{M}$ in order to match the XRISM observations. A spherical free-wind model can match the observed temperatures and densities using the $\dot{E}$ and $\dot{M}$ values from \cite{Boettcher} (as shown in that work), but the wind in M82 is almost certaintly nozzled to some extent. Once the opening angle is limited, none of the $\dot{E}$ and $\dot{M}$ values consistent with the temperatures and densities are sufficient to power the cool wind. 

The second thing to note is that the $\dot{E}$ value in \cite{Boettcher} is derived assuming that the $v_{\rm RMS}$ ($=\sqrt{3} \sigma_{\rm LOS,obs})$ value corresponds to the bulk outflow velocity. If we modify that assumption, as discussed in the next section, the derived $\dot{E}$ value would change. This also assumes that the outflow energetics are equivalent to the input energy and radiative losses are negligible, which we also discuss in the next section.

Finally, the CC-like models used in this work provide the simplest picture of a hot wind, and do not include much of the important physics that we know is at play in a superwind environment such as radiative cooling, cosmic rays, magnetic fields, and the interplay between different wind phases \citep[e.g.][]{Silich2004, thompson2016, gronke2020, bruggen2020, cottle2020, Ruszkowski2023,thompsonandheckman, lopez2025,Schneider2020}. Our simulations also do not include small scale motion that might contribute to the velocity broadening, and the $\dot{E}$ value that is used as an input for our simulation does not account for the extra kinetic energy that is implied from needing additional velocity broadening to fit the XRISM observations. Our best fit model with a $\sigma_{LOS} = 825$ km/s implies that the true energy in the driving region is higher than can be assumed using just the CC-like simulation input, but whether that energy goes into powering the outflow remains uncertain. A better understanding of these dynamics and the ways in which energy and material are exchanged will help us better understand the driving mechanism of M82's wind and the extent to which the supernova energy is thermalized. 

\subsection{Velocity Dispersion}
\label{sec:disc_vel}

XRISM allows for the first constraints on the velocity dispersion of the hot phase in M82, finding much higher velocities within the driving region of the starburst than predicted by free wind models such as those developed in \cite{CC}. However, interpreting these constraints is difficult given our inability to determine whether the dispersion comes from a bulk outflow or from smaller-scale motions driven by individual supernova remnants and turbulence. In \cite{Boettcher}, the observed velocity dispersion of $\sigma_{\rm LOS,obs}\approx$ 600 km/s was assumed to be dominated by the bulk outflow, and was interpreted to represent the outflow rate at the edge of the driving region. 

In this work, we examine another possibility. The CC model predicts a steep drop off in density at the edge of the driving region as seen in Figure \ref{fig:1dprofiles}, and the drop off is further supported by Chandra observations that limit the Fe XXV-bearing material to the core of the starburst as seen in Figure \ref{fig:chandra} \citep[e.g.][]{Strickland2007,Iwasawa2021}. This means that the emission is dominated by material in the driving region itself. Following the CC predictions, the bulk movement of the wind within this region is relatively small, increasing with radius but only reaching velocities of $v_{\rm RMS}\approx$600 km/s at the edge of the driving region, much slower than the $v_{\rm RMS, obs}\approx$1000 km/s observed by XRISM. 

Thus, we see that there is a significant inconsistency between the predictions of the free-wind model and the XRISM measurements of velocity dispersion. We cannot attribute this to uncertainties in the geometry, as we see across all opening angles considered significant additional broadening is needed to match the XRISM observations. It is possible that one major issue with the free-wind model is an underprediction of the density at larger radii, as the entrainment and heating of cool gas is not captured in this model. This could lead to lower-than-expected velocity dispersions from the mock spectra. However, we see from Chandra observations in Figure \ref{fig:chandra} that the Fe XXV emission is dominated by the central region.

This suggests that the observed velocity dispersion could be dominated by small-scale motions within the driving region not captured within a free wind model in which energy input is purely thermal. We could consider these motions to be caused by the expansion of supernova remnants and the generation of turbulent cascades within the interstellar medium (ISM), which motivates our pipeline of adding additional broadening to the spectra generated by a free-wind simulation. It is worth highlighting that this method is not meant to represent a self-consistent model for the wind driving of M82. Instead, we are wanting to understand the extent to which the velocity dispersion from a CC-like model is inconsistent with the XRISM observations, and consider the implications of a large $\sigma_{\rm LOS}$ driven by small-scale motions rather than bulk outflow. We are also constraining our analysis to within the driving region, which accounts for almost all of the Fe XXV emission arises, and we do not expect the velocity dispersion to be dominated by small-scale motions at larger radii.

One primary concern with interpreting the velocity dispersion as due to small-scale motions within the driving region is the impact this would have on the temperatures and overall energy budget of the starburst. The limits on our velocity dispersion of $\sigma_{\rm LOS} = 675-1245$ km/s correspond to 3D velocities of $\sqrt{3}\sigma_{\rm LOS}=v_{\rm RMS}$ = 1170-2156 km/s. We can determine the Mach number of this material by taking the average temperature weighted by density squared in our simulation, 1.77 keV. Our $v_{\rm RMS}$ values thus correspond to Mach 1.71-3.14. Solving for the postshock Mach number for planar shocks in a $\gamma = 5/3$ material gives us an expected maximum value of 1.3. This suggests that the temperature of the hot gas in M82 is much lower than we would expect given the observed velocity dispersion.

There are several possible explanations for this discrepancy. The first is that there might be a hotter gas phase than the $\approx$2 keV gas observed with XRISM, with temperatures more consistent with the observed velocities. However, the lack of an observed Fe XXVI line in the XRISM data places strong constraints on gas at this temperature, and our spectral fitting highly prefers models with temperatures closer to 2 keV in the driving region. Another possibility is that the central gas has undergone significant cooling, which is not included in our model.  This could occur through strong adiabatic expansion, although that is unexpected within the driving region of the starburst. Energy could also be radiated away, although the hard X-ray luminosity of $1.7 \times 10^{39}$ erg/s found in \cite{Iwasawa2023} is $\sim 0.7\%$ the input energy of our best fit fiducial model, suggesting such losses are negligible. 

One final possibility is that a significant fraction of the shock energy that we would expect to go into gas heating could instead go into amplifying magnetic fields and/or driving cosmic rays. 

We can estimate what the required energy budget for non-thermal sinks would be. To generate Mach 1.3 motion with our best fit $v_{\rm RMS}$ would require kT=$7.57$ keV. This represents a difference in thermal energy of $ 1.1 \times 10^{55}$ erg from our best fit model. If we assume this energy is going into magnetic fields, that would result in field strengths of $\approx$ 530$\mu$G. This is very similar to the field strengths we would get if we assume magnetic pressure is equal to the thermal pressure, B $\approx$ 500$\mu$G. This is probably an upper limit, as while some works suggest equipartition in environments where high-velocity supernova-driven turbulence dominates \citep{lacki2013}, other simulations of magnetized turbulent media predict $\beta>1$ \citep{Sur2014}. Recent observations have found B$\approx$ 100-500 $\mu$G in the central region of M82 \citep[e.g.][]{Buckman2020,lopez2021, Persic2025}, which would suggest that strong magnetic fields could account for some of the energy loss. Significant energy going into cosmic rays is also not unexpected, as it has been postulated that cosmic rays could play a major role in driving the superwind \citep[e.g.][]{Ipavich1975,Zweibel2017, Quataert2022, Ruszkowski2023}. 

Understanding how the discrepancy between the model and observations is driven by geometry, the lack of turbulent motions, and the non-inclusion of physics such as cooling and magnetic fields will require more detailed simulation work going forward.

It is also worth discussing the discrepancy between our best fit velocity dispersion of $\sigma_{\rm LOS}$ = 825$^{+420}_{-150}$ km/s and that found in \cite{Boettcher} of $\sigma_{\rm LOS} = 595^{+464}_{-128}$ km/s. This difference can be attributed to two things: the model and the fitting region. \cite{Boettcher} uses an isothermal model, whereas our simulations display a more complex temperature structure as shown in Figure \ref{fig:1dprofiles}. Additionally, while \cite{Boettcher} fits over the entire spectrum, we choose to focus on smaller regions centered around the S XVI and Fe XXV regions. This choice was made due to our simulations being designed to only model the hottest gas phases, so fitting the entire spectra would rely on the inclusion of other components that would introduce more free parameters. It is also worth noting that despite the lower value, the best fit from \cite{Boettcher} still represents a supersonic solution if interpreted as non-uniform motion within the driving region. 

\section{Summary}

In this work, we provide the first in-depth test of a CC-like free wind model using high-resolution X-ray spectroscopy of the starburst galaxy M82 taken with XRISM. We generate a suite of hydrodynamic simulations parameterized by energy and mass input rates with three different opening angles, create mock XRISM-Resolve spectra, and fit against the XRISM observations. As our free-wind model alone is unable to match the observed velocity dispersion, we add additional velocity broadening to the model and fit for $\sigma_{\rm LOS}$. Our main findings are as follows: 

\begin{enumerate}

\item For a starburst with a spherical driving region with a radius of 200 pc and a half opening angle of 30$\degree$, we can place constraints on our model of $\dot{M}$ = 0.52$^{+0.020}_{-0.021}$ $M_\odot$/yr, $\dot{E}$ = 2.4$^{+0.1}_{-0.09}\times 10^{41}$ erg/s and $\sigma_{\rm LOS}$ = 825$^{+420}_{-150}$ km/s.

\item As the geometry of M82 is uncertain, we can set upper and lower limits on energy and mass loading for half opening angles of 15$\degree$-60$\degree$, finding $\dot{M} \approx 0.1-1.6$ $M_\odot$/yr and $\dot{E} \approx 0.5-8 \times 10^{41}$ erg/s. 

\item These best-fit energetic constraints are insufficient to power the cooler phases of the outflow, although it is worth noting that there are significant uncertainties in the geometry, which could change these results.

\item The XRISM results provide the first measure of the hot phase velocity dispersion, which we find the free-wind model unable to replicate across a range of geometries and energies. We interpret this to mean that there are additional disorganized nonthermal motions within the driving region that are unaccounted for in the free-wind model, likely due to the evolution of individual supernova remnants and turbulent cascades in the ISM. This material would need to be moving at a Mach number of approximately 2 to be consistent with the observations. This Mach number is inconsistent with the properties of nonradiative, hydrodynamic shocks which would lead to hotter material, but a significant fraction of the shock energy could go into cosmic rays and magnetic field amplification.
\end{enumerate}

Overall, we find that a free-wind model alone is inconsistent with the XRISM observations. Our approach of adding additional velocity broadening to a free-wind-like model is successful in matching high-resolution spectra of the hot phase of M82's wind. However, this is not a self-consistent model, and the high velocity dispersion observed by XRISM presents a puzzle for free-wind models, as it represents supersonic motion happening within the driving region of the starburst. More modeling of the role of magnetic fields and cosmic rays will be key to understanding how such velocities are possible. Spatially resolved measurements of the velocity structure of the hot wind will also prove useful, particularly when it comes to disentangling the bulk outflow from smaller-scale motions.

\section*{Acknowledgements}

We would like to thank the reviewer for their helpful comments that greatly improved this paper. We would also like to thank Peter Smith, Seth Cohen, Darby Kramer, Brad Koplitz, and the XRISM M82 Working Group for their helpful discussions during the development of this work. S.G. acknowledges support from NSF Grant No. 2233001.  E.S. acknowledges support from NASA grants 80NSSC22K1265 and  80NSSC23K0646.  P.G. acknowledges funding by the Deutsche Forschungsgemeinschaft (DFG) – 555983577. M.B. acknowledges support from the DFG under Germany's Excellence Strategy - EXC 2121 "Quantum Universe" - 390833306 and from the BMBF ErUM-Pro grant 05A2023. EB acknowledges support by NASA under award number 80GSFC24M0006. The authors also gratefully acknowledge the Gauss Centre for Supercomputing e.V. (www.gauss-centre.eu) for providing computing time through the John von Neumann Institute for Computing (NIC) on the GCS Supercomputer JUWELS at J\"ulich Supercomputing Centre (JSC).

\vspace{5mm}

\newpage

\bibliographystyle{aasjournal}
\bibliography{bib.bib}

\end{document}